\documentclass[10pt,a4paper]{article}

\usepackage[utf8]{inputenc} 
\usepackage[T1]{fontenc}    %
\usepackage[english]{babel} 
\usepackage[final]{microtype} 

\usepackage{amssymb,amsmath,mathtools} 
\usepackage{graphicx} 
\usepackage{bbm} 
\usepackage{tikz} 

\usepackage[font=small,labelfont=bf]{caption}

\usepackage[hmargin=0.12\paperwidth,vmargin=0.2\paperwidth,bindingoffset=0.5cm]{geometry}

\usepackage{titlesec}

\titleformat*{\section}{\large\bfseries}
\titleformat*{\subsection}{\bfseries}

\newcommand{\p}{\partial} 

\newcommand{\N}{\mathbb{N}} 

\newcommand{\R}{\mathbb{R}} 
\newcommand{\C}{\mathbb{C}}

\renewcommand{\phi}{\varphi} 
\renewcommand{\epsilon}{\varepsilon} 

\DeclareMathOperator*{\perm}{per} 
\DeclareMathOperator{\sign}{sign} 
\DeclareMathOperator{\erfc}{erfc} 

\DeclarePairedDelimiter{\abs}{\lvert}{\rvert} 
\DeclarePairedDelimiter{\average}{\langle}{\rangle} 

\newcommand{\MeijerG}[8][\bigg]{G^{{ #2 },{ #3 }}_{{ #4 },{ #5 }} #1( \begin{matrix} #6 \\ #7 \end{matrix}\, #1\vert\, #8 #1)}

\usepackage{hyperref}   
\hypersetup{
colorlinks=true,        
linktoc=page,           
linkcolor=blue,         
citecolor=blue,         
}

\title{\bfseries Lyapunov exponents for products of rectangular\\ real, complex and quaternionic Ginibre matrices}
\author{J.~R.~Ipsen\\
\small Department of Physics, Bielefeld University,\\
\small Postfach 100131, D-33501 Bielefeld, Germany}
\date{\today}

\begin{document}	

\maketitle

\begin{abstract}
\noindent
We study the joint density of eigenvalues for products of independent rectangular real, complex and quaternionic Ginibre matrices. In the limit where the number of matrices tends to infinity, it is shown that the joint probability density function for the eigenvalues forms a permanental point process for all three classes. The moduli of the eigenvalues become uncorrelated and log-normal distributed, while the distribution for the phases of the eigenvalues depends on whether real, complex or quaternionic Ginibre matrices are considered. In the derivation for a product of real matrices, we explicitly use the fact that all eigenvalues become real when the number of matrices tends to infinity. Finally, we compare our results with known results for the Lyapunov exponents as well as numerical simulations.

\end{abstract}

\section{Introduction}
\label{sec:intro}

It is well-known that products of random matrices occur as a natural tool in many areas of statistical physics, for example, the transfer matrix description of disordered magnetic systems and electric transport properties of disordered materials, see~\cite{CPV:1993,Beenakker:1997,Haake:2010} and references within. Other active research areas are also related to products of random matrices, such as quantum chromodynamics at finite chemical potential~\cite{Osborn:2004,Akemann:2007}, ecological stability~\cite{Caswell:2001}, wireless telecommunication~\cite{Muller:2002,TV:2004}, and free probability~\cite{NS:2006}. 

Many applications are concerned with asymptotic properties in the limit where the number of matrices tends to infinity. Some of the most important quantities in this limit are the Lyapunov exponents. As an example it could be mentioned that products of random matrices can be used to mimic chaoticity in dynamical systems, see e.g.~\cite{Benettin:1984,PV:1986}. Here the Lyapunov exponents give the inverse time rate by which nearby trajectories diverge. A further example is the disordered Ising chain which can be described using a product of random transfer matrices~\cite{DVP:1978}; the free energy density is then given by the largest Lyapunov exponent.

There exist a few general theorems regarding Lyapunov exponents. We recall the Furstenberg--Kesten theorem for the largest Lyapunov exponent~\cite{FK:1960} and Oseledec's multiplicative ergodic theorem~\cite{Oseledec:1968,Raghunathan:1979}. However, usually it is very difficult to obtain analytical expressions for the Lyapunov exponents, and explicit formulae are only known for a few simple models. These models typically involve small matrices (most often $2\times 2$ matrices) or randomness described by a single random parameter, see e.g.~\cite{JLJN:2002}. An exception to this was presented in~\cite{Newman:1986}, which investigates the Lyapunov exponents for a product of real Ginibre matrices. This result was extended to complex and quaternionic Ginibre matrices with a deterministic correlation matrix~\cite{Forrester:2013,Kargin:2014}.

Recently, it has been shown that the joint probability density function for both the eigenvalues~\cite{AB:2012} and singular values~\cite{AKW:2013,AIK:2013} form determinantal point processes for a finite product of complex Ginibre matrices. This has sparked new interest in products of Ginibre matrices and related models. In particular, it was shown that the joint probability density function for the eigenvalues of real and quaternionic Ginibre matrices form Pfaffian point processes~\cite{Ipsen:2013,Forrester:2014a,IK:2014}. 

At finite matrix dimensions, a finite product of real Ginibre matrices is the most complicated situation. In this case the eigenvalue spectrum consists of real eigenvalues as well as complex conjugate pairs, and an explicit form of the joint probability density function is only known for a single matrix~\cite{LS:1991,Edelman:1997} and for the product two matrices~\cite{APS:2009,APS:2010} (a representation in terms of $2\times 2$ matrices is known for an arbitrary number of matrices~\cite{IK:2014}). The situation simplifies when the matrix dimensions are kept fixed while the number of matrices is taken to infinity. In this limit, the fraction of real eigenvalues tends to unity, which is extremely useful since an explicit expression for the joint probability density function is known in this case. This phenomenon was first observed in~\cite{Lakshminarayan:2013} and later proved for arbitrary matrix dimension in~\cite{Forrester:2014a}.

\begin{figure}[htbp]
\center
\includegraphics{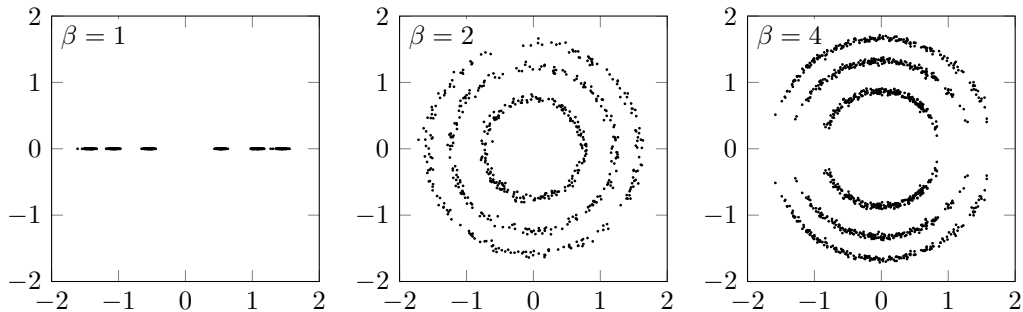}
\caption{Each scatter plot shows eigenvalues of $200$ realizations of a product matrix constructed as a product of $200$ real (left panel), complex (center panel), or quaternionic (right panel) $3\times 3$ Ginibre matrices; the eigenvalues have been rescaled by taking the $1/200$-th power of the moduli. For a single realization of a product matrix each ring contains exactly one eigenvalue ($\beta=1,2$) or exactly one complex conjugate pair ($\beta=4$). Note that \emph{all} the eigenvalues are real for $\beta=1$, while the eigenvalues are scattered in the complex plane for $\beta=2$ and $\beta=4$.}
\label{fig:scatter}
\end{figure}

In~\cite{ABK:2014} the joint probability density function for eigen- and singular values for a finite product of complex Ginibre matrices were linked to known results for the Lyapunov exponents~\cite{Newman:1986,Forrester:2013}. In this paper, we extend this discussion to the statistical properties of the eigenvalues of a product of independent real and quaternionic Ginibre matrices. Figure~\ref{fig:scatter} shows the eigenvalue distributions for products of real ($\beta=1$), complex ($\beta=2$), and quaternionic ($\beta=4$) Ginibre matrices and illustrates the problem in question. For a product of real Ginibre matrices the eigenvalues are clustered within narrow and separated bands; for products of complex and quaternionic Ginibre matrices the eigenvalues are distributed along narrow and well-separated rings. Looking at figure~\ref{fig:scatter} several natural questions arise: (i) What are the radii of the rings? (ii) What is the radial distribution in the neighbourhood of the rings? (iii) What is the 
angular distribution within the rings? (iv) And how are the eigenvalues correlated? We will provide explicit answers to the all of these questions. Note that the statistical 
properties of the eigenvalues for a product of square, complex Ginibre matrices was recently described in~\cite{ABK:2014}. Our results generalize this result by considering rectangular matrices and including results for both products of real ($\beta=1$) and quaternionic ($\beta=4$) Ginibre matrices. In particular, we confirm a conjecture made in~\cite{ABK:2014} regarding the angular dependence for real and quaternionic matrices.

The paper is organized as follows: In section~\ref{sec:results}, we summarize the background, state our main results and compare them with numerical data. Derivations of the results for complex, quaternionic and real matrices are presented in section~\ref{sec:complex},~\ref{sec:quaternion} and~\ref{sec:real}, respectively. Conclusions and an outlook are given in section~\ref{sec:conclusion} while some identities of the Meijer $G$-function are given in an appendix for easy reference.

\section{Background and main statements}
\label{sec:results}

Let us consider a dynamical system in the variables $u(t)=\{u_1(t),\ldots,u_N(t)\}$ evolving in discrete time according to the law $u_i(t+1)=f_i[u(t)]$, where each $f_i$ is some smooth function. Such systems arise in a vast variety of contexts in the physical, biological and social sciences. In order to study the separation between two nearby trajectories, we introduce the tangent vector $\delta u(t)$ which may be regarded as a small perturbation. This gives rise to the linearised problem
\begin{equation}
\delta u(t+1)=X_t\,\delta u(t) \qquad\text{with}\qquad (X_t)_{ij}\equiv \frac{\p f_i(u)}{\p u_j}\bigg\vert_{u(t)},
\end{equation}
where $X_t$ is known as the stability matrix. The evolution of the system is described by a product matrix,
\begin{equation}
Y(t)\equiv X_tX_{t-1}\cdots X_1,\qquad t\in\N,
\label{results:product}
\end{equation}
which is obtained by evaluating the stability matrix along the trajectory.

The simplest examples occur when we consider evolution in the vicinity of a fixed point. In these cases the stability matrices $X_i$ ($i=1,\ldots,t$) become delta-correlated and we have $Y(t)=(X_1)^t$. For many large complex systems a reasonable approximation can be obtained by choosing $X_1$ as a random matrix with symmetries determined by the underlying theory~\cite{GA:1970}. An example is provided by the evolution of large ecosystems where minimal requirements demand that the stability matrix is an asymmetric real matrix~\cite{May:1972}.

More challenging problems occur when studying evolution away from the fixed points. A first non-trivial approximation is provided by choosing the stability matrices $X_i$ ($i=1,\ldots,t$) as independent random matrices~\cite{Benettin:1984,PV:1986}, see also~\cite{CPV:1993}. This approximation is reasonable in the limit of strong chaoticity where the stability matrices are only weakly correlated.

In this paper we study one of the simplest toy models for such evolutions. The stability matrices $X_i$ ($i=1,\ldots,t$) are assumed to be independent $(N+\nu_i)\times (N+\nu_{i-1})$ rectangular Ginibre matrices, i.e. matrices where the entries are independent centred Gaussian random variables. Here $N$ is a positive integer denoting the smallest matrix dimension, while $\nu_i$ are non-negative integers incorporating the rectangular structure. It turns out that the ordering of the matrix dimensions is irrelevant for the (non-zero) eigen- and singular values. This is a consequence of a weak commutation relation for isometric random matrices~\cite{IK:2014}. For this reason we can choose the ordering $0=\nu_0\leq \nu_1\leq \cdots \leq\nu_t$ without loss of generality; for simplicity we will also assume that there is an upper bound for the rectangularity $\nu_t\to\nu_\infty<\infty$. We use the standard classification and consider three distinct classes: real ($\beta=1$), complex ($\beta=2$), and quaternionic ($\beta=4$) matrices. For 
quaternions we use the canonical representation as 
$2\times 2$ matrices, see e.g.~\cite{Mehta:2004}. Thus, an $N\times M$ quaternionic matrix should be understood as a $2N\times 2M$ complex matrix which satisfies the quaternionic symmetry requirements.

These Gaussian models are particular interesting since they are exactly solvable for any values of the above given parameters. The joint probability density function for the eigenvalues of the product matrix~\eqref{results:product} with dimension $N$ and Dyson index $\beta$ at time $t$ is~\cite{AB:2012,Ipsen:2013,ARRS:2013,Forrester:2014a,IK:2014},
\begin{subequations}
\label{results:jpdf-all}
\begin{align}
P_{N}^{\beta=1}(x_1,\ldots,x_N;t)&=\frac{1}{Z_{N,\nu}^{\beta=1}}\prod_{1\leq k<\ell\leq N}\abs{x_k-x_\ell}\prod_{n=1}^N w_{\nu}^{\beta=1}(x_n), \label{results:jpdf-real} \\
P_{N}^{\beta=2}(z_1,\ldots,z_N;t)&=\frac{1}{Z_{N,\nu}^{\beta=2}}\prod_{1\leq k<\ell\leq N}\abs{z_k-z_\ell}^2\prod_{n=1}^N w_{\nu}^{\beta=2}(z_n), \label{results:jpdf-complex} \\
P_{N}^{\beta=4}(z_1,\ldots,z_N;t)&=\frac{1}{Z_{N,\nu}^{\beta=4}}\prod_{1\leq k<\ell\leq N}\abs{z_k-z_\ell}^2\abs{z_k-z_\ell^*}^2\prod_{n=1}^N\abs{z_n-z_n^*}^2 w_{\nu}^{\beta=4}(z_n),
\label{results:jpdf-quaternion}
\end{align}
\end{subequations}
where $Z_{N,\nu}^{\beta}$ are constants and $w_{\nu}^{\beta}(z)$ are the weight functions; both will be specified below.
For $\beta=2$ the eigenvalues can lie anywhere in the complex plane, $z_n\in\C$; for $\beta=4$ the eigenvalues come in complex conjugate pairs and $z_n$ denotes the value in the upper half of the complex plane, $z_n\in\C_+$. The situation for $\beta=1$ is more complicated. Rather than considering the full joint probability density, we restrict ourselves to the case where all eigenvalues are real, $x_n\in \R$. In fact, this is \emph{not} a restriction in the large time limit for finite size matrices, since in this limit all eigenvalues are real almost surely~\cite{Forrester:2014a,Lakshminarayan:2013}. This will be reaffirmed by the explicit calculations in this paper. The constants, $Z_{N,\nu}^\beta$, depend on the matrix dimensions through $N$ and the integers $\nu_i$ ($i=1,\ldots,t$) collectively denoted by $\nu$. Explicitly, we have
\begin{subequations}
\label{results:normal}
\begin{align}
Z_{N,\nu}^{\beta=1}&=N!\,2^{\,tN(N+1)/4}\prod_{n=1}^N\prod_{i=1}^t\Gamma\Big[\frac{\nu_i}{2}+\frac{n}{2}\Big], & 
\lim_{t\to\infty}\prod_{n=1}^N\int_{\R}dx_n\,P_{N}^{\beta=1}(x_1,\ldots,x_N;t)&=1, \label{results:normal-real} \\
Z_{N,\nu}^{\beta=2}&=N!\,\pi^N\prod_{n=1}^N\prod_{i=1}^t\Gamma[\nu_i+n], & 
\prod_{n=1}^N\int_{\C}d^2z_n\,P_{N}^{\beta=2}(z_1,\ldots,z_N;t)&=1, \label{results:normal-complex} \\
Z_{N,\nu}^{\beta=4}&=\frac{N!\,\pi^N}{2^{\,tN(N+1)}}\prod_{n=1}^N\prod_{i=1}^t\Gamma[2\nu_i+2n], & 
\prod_{n=1}^N\int_{\C_+}d^2z_n\,P_{N}^{\beta=4}(z_1,\ldots,z_N;t)&=1. \label{results:normal-quaternion}
\end{align}
\end{subequations}
The constants $Z_{N,\nu}^{\beta=2}$ and $Z_{N,\nu}^{\beta=4}$ are normalization constants, such that integration over the eigenvalues yields unity. In the real case, we have chosen $Z_{N,\nu}^{\beta=1}$ such that integration over the eigenvalues yields the probability that all eigenvalues are real. At small $t$ this probability will be less than one, but it is equal to unity in the large-$t$ regime that we consider in this paper (as we shall see). The weight functions in~\eqref{results:jpdf-all} can collectively be expressed in terms of a Meijer $G$-function (see appendix~\ref{sec:meijer}),
\begin{equation}
w_{\nu}^{\beta}(z)=\MeijerG{t}{0}{0}{t}{-}{\frac{\beta\nu_1}{2},\ldots,\frac{\beta\nu_t}{2}}{\Big(\frac{\beta}{2}\Big)^t\,\abs z^2}.
\label{results:weight}
\end{equation}
We stress that the joint probability density functions~\eqref{results:jpdf-all} are valid at any time. However, in this paper we focus on the large time limit. 

Let us first recall the usual definition of the Lyapunov exponents for products of random matrices.
If $\sigma_n(t)$ ($n=1,\ldots,N$) denote the singular values of the product matrix~\eqref{results:product}, then the Lyapunov exponents are defined by
\begin{equation}
\mu_n\equiv \lim_{t\to\infty}\frac{\log \sigma_n(t)}{t}.
\label{results:def-lyaponov-singular}
\end{equation}
In fact, the Lyapunov exponents are explicitly known for a product of Ginibre matrices~\cite{Newman:1986,Forrester:2013,Kargin:2014}. Up to a reordering, we have
\begin{equation}
\mu_n^\beta=\frac{1}{2}\log\frac{2}{\beta}+\frac{1}{2}\average[\Big]{\psi\Big(\frac{\beta (\nu+n)}{2}\Big)},
\label{results:lyaponov}
\end{equation}
where $\psi(x)\equiv \Gamma'[x]/\Gamma[x]$ is the digamma function and $\beta$ is the Dyson index specifying whether we consider real ($\beta=1$), complex ($\beta=2$), or quaternionic ($\beta=4$) matrices. The time average is defined as
\begin{equation}
\average[\Big]{\psi\Big(\frac{\beta (\nu+n)}{2}\Big)}\equiv \lim_{t\to\infty} \frac1t\sum_{i=1}^t \psi\Big(\frac{\beta (\nu_i+n)}{2}\Big).
\label{results:average}
\end{equation}
Recall that $n=1,\ldots,N$ and $\nu_t\to\nu_\infty<\infty$ which ensures that the limit is well-defined without further rescaling. Note that if we consider square matrices ($\nu_i=0$ for all $i$) then the average becomes trivial, $\average{\psi(\beta n/2)}=\psi(\beta n/2)$.

Not only Lyapunov exponents themselves but also their fluctuations are essential for the description of the large time behaviour of~\eqref{results:product}. For a chaotic system, where nearby trajectories diverge, the fluctuations will increase compared with length scale of the system even though they decrease compared to the interspacing between Lyapunov exponents. For this reason, we are not only interested the strict $t\to\infty$ limit of the Lyapunov exponents, but also their analogous at large but finite $t$. Given definition~\eqref{results:def-lyaponov-singular}, it would be natural to define the Lyapunov exponents at finite times by
\begin{equation}
\gamma_n(t)\equiv \frac{\log \sigma_n(t)}{t}.
\label{results:def-lyaponov-finite-singular}
\end{equation}
However, in this paper we will use a different definition of the finite-time Lyapunov exponents. We will use the eigenvalues of the product matrix~\eqref{results:product} to define the Lyapunov exponents rather than the singular values. If $z_n(t)$ ($n=1,\ldots,N$) denote the eigenvalues of the product matrix, then we define the finite-time Lyapunov exponents by
\begin{equation}
\lambda_n(t)\equiv \frac{\log \abs{z_n(t)}}{t},
\label{results:def-lyaponov-finite}
\end{equation}
where $\abs{z_n(t)}$ are the moduli of the eigenvalues. The exponents~\eqref{results:def-lyaponov-finite} are also known as stability exponents~\cite{GSO:1987}. Definition~\eqref{results:def-lyaponov-finite-singular} and~\eqref{results:def-lyaponov-finite} are equivalent in the sense that they converge to same limit as $t$ tends to infinity, namely to~\eqref{results:lyaponov}. A priori this equivalence is far from obvious. In particular, we know that the eigen- and the singular values behave very different at small $t$. 

A benefit of using definition~\eqref{results:def-lyaponov-finite} rather than~\eqref{results:def-lyaponov-finite-singular} is that explicit expressions for the joint density of the eigenvalues are known for all three Dyson indices~\eqref{results:jpdf-all}, while the joint density for the singular values is only explicitly known for $\beta=2$, see~\cite{AKW:2013,AIK:2013}. Note that definition~\eqref{results:def-lyaponov-finite} is related to the eigenvalues through the parametrization  $z_n=e^{t\lambda_n+i\theta_n}$, where $e^{t\lambda_n}$ is the modulus and $\theta_n$ is the phase of the eigenvalue. For $\beta=2$ the eigenvalues can take any complex value, hence $\theta_n\in [0,2\pi)$. For $\beta=1$ the eigenvalues are assumed to be real (which we emphasize by writing $z_n=x_n$) and it follows that $\theta_n\in\{0,\pi\}$. For $\beta=4$ the eigenvalues come in complex conjugate pairs and $z_n$ denote the values in the upper half of the complex plane, hence $\theta_n\in[0,\pi]$. Figure~\ref{fig:scatter} 
shows scatter plots of the eigenvalues, $e^{\lambda_n+i\theta_n}$. Moreover, the eigenbasis provide an alternative (non-orthogonal) basis which generally differ from the Lyapunov basis. This basis is known as the stability basis and has important geometrical interpretation, since any perturbation $\delta u$ is projected onto the subspace spanned by the eigenvectors corresponding the largest stability exponent exponentially fast.

We can now formulate the main statement of this paper. We introduce the joint density of the finite-time Lyapunov exponents,
\begin{equation}
\rho_N^\beta(\lambda_1,\theta_1,\ldots,\lambda_N,\theta_N;t)\equiv \prod_{n=1}^Nte^{\gamma t\lambda_n}\,P_N^\beta(e^{t\lambda_1+i\theta_1},\ldots,e^{t\lambda_N+i\theta_N};t),
\end{equation}
where $P_N^\beta(z_1,\ldots,z_N;t)$ is given by~\eqref{results:jpdf-all} and $\gamma=1$ for $\beta=1$ and $\gamma=2$ for $\beta=2,4$. The prefactor on the right hand side is due to the change of variables. At large times ($t\gg1$) the mutual interaction between eigenvalues vanish and the joint densities turn into permanental point processes,
\begin{subequations}
\label{results:permanent}
\begin{align}
\rho_N^{\beta=1}(\lambda_1,\theta_1,\ldots,\lambda_N,\theta_N;t\gg1)&=\frac{1}{N!}\perm_{1\leq k,\ell\leq N}\bigg[\frac{1}{2}f_k^{\beta=1}(\lambda_\ell;t\gg1) \bigg]\label{results:permanent-1}, \\
\rho_N^{\beta=2}(\lambda_1,\theta_1,\ldots,\lambda_N,\theta_N;t\gg1)&=\frac{1}{N!}\perm_{1\leq k,\ell\leq N}\bigg[\frac{1}{2\pi}f_k^{\beta=2}(\lambda_\ell;t\gg1) \bigg]\label{results:permanent-2}, \\
\rho_N^{\beta=4}(\lambda_1,\theta_1,\ldots,\lambda_N,\theta_N;t\gg1)&=\frac{1}{N!}\perm_{1\leq k,\ell\leq N}\bigg[\frac{2\sin^2\theta_\ell }{\pi}f_k^{\beta=4}(\lambda_\ell;t\gg1) \bigg]
\label{results:permanent-4}
\end{align}
\end{subequations}
for real, complex and quaternionic matrices, respectively. Here ``$\perm$'' denotes the permanent and
\begin{equation}
f_{n}^{\beta}(\lambda;t\gg 1)=\frac{1}{\sqrt{2\pi} \sigma_{n}^{\beta}}\exp\bigg[-\frac{(\lambda-\mu_{n}^{\beta})^2}{2(\sigma_{n}^{\beta})^2}\bigg],
\end{equation}
where the mean is given by equation~\eqref{results:lyaponov} and the variance is given by
\begin{equation}
(\sigma_n^\beta)^2=\frac{1}{4t}\average[\Big]{\psi'\Big(\frac{\beta (\nu+n)}{2}\Big)}.
\label{results:variance}
\end{equation}
The average in~\eqref{results:variance} is defined similar to~\eqref{results:average} and $\psi'(x)$ denotes the first derivative of the digamma function also known as the trigamma function. The derivation of the permanental structure~\eqref{results:permanent} is given in section~\ref{sec:complex},~\ref{sec:quaternion} and~\ref{sec:real}.


We recall that $\theta_n\in\{0,\pi\}$, $\theta_n\in[0,2\pi)$, and $\theta_n\in[0,\pi]$ for real, complex and quaternionic matrices, respectively. Integrating out the phases, we obtain the joint density for the Lyapunov exponents only:
\begin{subequations}
\label{results:lyaponov-permanent}
\begin{align}
\prod_{n=1}^N\sum_{\theta_n=0,\pi}\rho_N^{\beta=1}(\lambda_1,\theta_1,\ldots,\lambda_N,\theta_N;t\gg1)&=\frac{1}{N!}\perm_{1\leq k,\ell\leq N}\big[f_k^{\beta=1}(\lambda_\ell;t\gg1) \big], \\
\prod_{n=1}^N\int_0^{2\pi} \!\!d\theta_n\,\rho_N^{\beta=2}(\lambda_1,\theta_1,\ldots,\lambda_N,\theta_N;t\gg1)&=\frac{1}{N!}\perm_{1\leq k,\ell\leq N}\big[f_k^{\beta=2}(\lambda_\ell;t\gg1) \big], \\
\prod_{n=1}^N\int_0^{\pi} \!d\theta_n\,\rho_N^{\beta=4}(\lambda_1,\theta_1,\ldots,\lambda_N,\theta_N;t\gg1)&=\frac{1}{N!}\perm_{1\leq k,\ell\leq N}\big[f_k^{\beta=4}(\lambda_\ell;t\gg1) \big].
\end{align}
\end{subequations}
Hence the Lyapunov exponents for all three classes obtain a similar structure in the large time limit. Furthermore, if we take the strict limit $t\to\infty$, then the variances tend to zero and the Gaussians turn into delta peaks, 
\begin{equation}
f_n^\beta(\lambda,t)\longrightarrow\delta(\lambda-\mu_n^\beta).
\end{equation}
In words this means that at large times the Lyapunov exponents become independent Gaussian random variables converging to a deterministic limit. We see that the deterministic values of Lyapunov exponents are given by~\eqref{results:lyaponov} even though the derivation is based on~\eqref{results:def-lyaponov-finite} rather than the canonical definition~\eqref{results:def-lyaponov-singular}.

\begin{figure}[htbp]
\center
\includegraphics{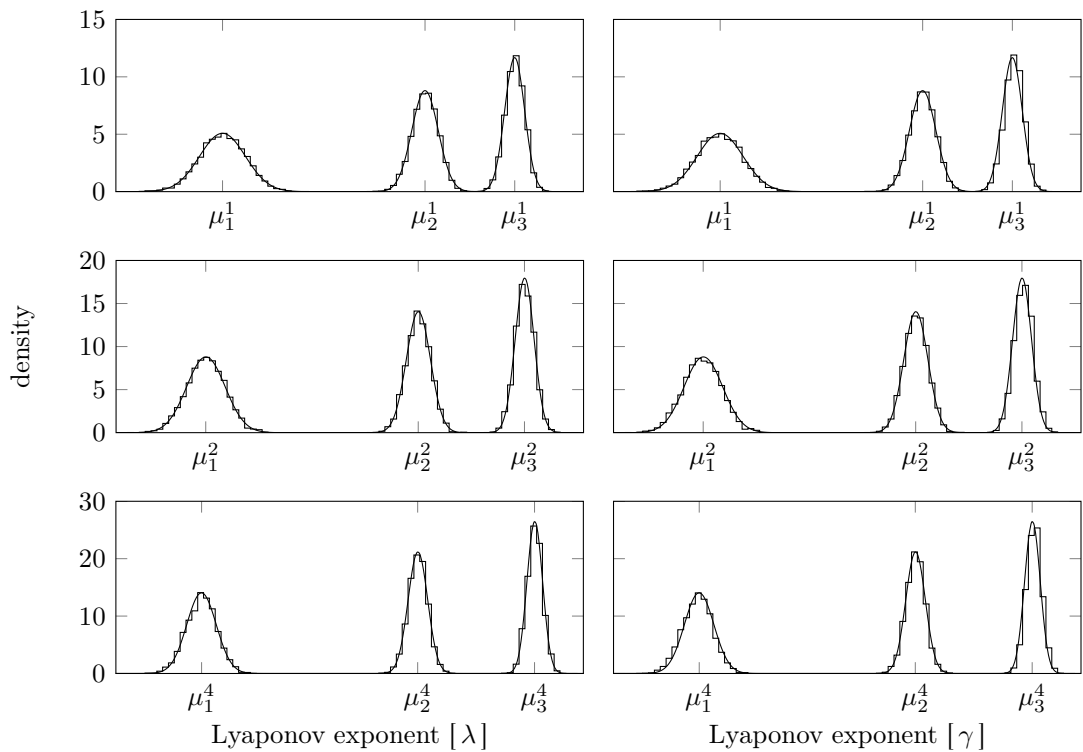}
\caption{Comparison of numerical simulations and analytical expressions for the density of the Lyapunov exponents at large times. The histograms show Lyapunov exponents for $4\,000$ realizations for a product of independent real (top row), complex (middle row), and quaternionic (bottom row) $3\times 3$ Ginibre matrices at time $t=200$. The Lyapunov exponents are constructed both from the eigenvalues (left column) and the singular values (right column). The solid curves show the analytical predictions~\eqref{results:lyaponov-permanent}. Note that the scales of both axes are different in the three different rows.}
\label{fig:histogram}
\end{figure}

In figure~\ref{fig:histogram}, we compare the analytical expression~\eqref{results:lyaponov-permanent} for the Lyapunov exponents at large times with numerical data constructed both from the eigenvalues~\eqref{results:def-lyaponov-finite} and from the singular values~\eqref{results:def-lyaponov-finite-singular}. Albeit the derivation of~\eqref{results:lyaponov-permanent} is based on the definition~\eqref{results:def-lyaponov-finite}, the expressions are in perfect agreement with numerical data for the singular values as well. For complex ($\beta=2$) matrices it was shown that the joint density of the Lyapunov exponents constructed through~\eqref{results:def-lyaponov-finite-singular} and~\eqref{results:def-lyaponov-finite} are identical at large times~\cite{ABK:2014}. This equivalence  holds for real ($\beta=1$) and quaternionic ($\beta=4$) matrices as well (compare the variances found in this paper with those presented in~\cite{Forrester:2015}). This explains the agreement seen in figure~\ref{fig:histogram}. In fact, the equivalence between the Lyapunov exponents~\eqref{results:def-lyaponov-finite-singular} and the stability exponents~\eqref{results:def-lyaponov-finite} at large times is conjectured to hold whenever the spectrum is non-degenerate~\cite{GSO:1987}.
\begin{figure}[htbp]
\center
\includegraphics{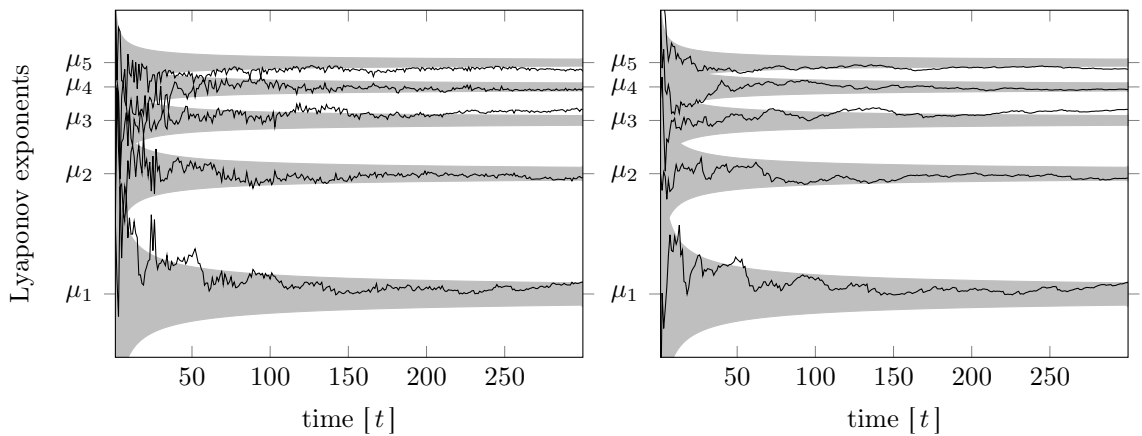}
\caption{A single realization of the Lyapunov exponents for a product of real $5\times 5$ Ginibre matrices as a function of time, $t$. The Lyapunov exponents are generated both from the eigenvalues (left panel) and from the singular values (right panel). The grey areas show the most likely positions predicted by the asymptotic approximation, $\mu_n^{1}\pm \sigma_n^{1}$ given by~\eqref{results:lyaponov} and~\eqref{results:variance}.}
\label{fig:convergence}
\end{figure}

Figure~\ref{fig:convergence} shows the convergence of Lyapunov exponents as a function of time constructed from both the eigenvalues~\eqref{results:def-lyaponov-finite} and the singular values~\eqref{results:def-lyaponov-finite-singular}. It should be noted that the variance~\eqref{results:variance} provide the order of magnitude of the sample-to-sample fluctuations, i.e. fluctuations from one realization of the product matrix~\eqref{results:product} to another. As mentioned above, the sample-to-sample fluctuation are the same for~\eqref{results:def-lyaponov-finite} and~\eqref{results:def-lyaponov-finite-singular}. We have not made any prediction for the fluctuations between successive times $t$ and $t+1$, but figure~\ref{fig:convergence} suggests that the time-to-time fluctuations of singular values~\eqref{results:def-lyaponov-finite-singular} are smaller than those of the eigenvalues~\eqref{results:def-lyaponov-finite}.

\section{Derivation for complex matrices}
\label{sec:complex}

In this section we will show that the joint probability density function for the eigenvalues of a product of complex Ginibre matrices~\eqref{results:jpdf-complex} leads to a permanental point process for the Lyapunov exponents in the large time limit, $t\gg1$. This has already been shown for a product of square matrices in~\cite{ABK:2014} and the generalization to rectangular matrices is straightforward. However, we find it enlightening to go through the derivation for products of complex matrices before we go on to discuss quaternionic and real matrices, since these derivations will have certain similarities with the complex case.

We closely follow the steps in~\cite{ABK:2014} and start by changing variables from the eigenvalues, $z_n$, to the Lyapunov exponents, $\lambda_n$, and the phases, $\theta_n$. The joint density~\eqref{results:jpdf-complex} becomes
\begin{equation}
\rho_{N}^{\beta=2}(\lambda_1,\theta_1,\ldots,\lambda_N,\theta_N;t)=
\frac{1}{Z_{N,\nu}^{\beta=2}}\prod_{1\leq k<\ell\leq N}\abs*{e^{t\lambda_k+i\theta_k}-e^{t\lambda_\ell+i\theta_\ell}}^2\prod_{n=1}^N te^{2t\lambda_n} w_{\nu}^{\beta=2}(e^{t\lambda_n}).
\label{complex:jpdf-cov}
\end{equation}
Using the standard form of a Vandermonde determinant, $\det_{k,\ell}[x_\ell^{k-1}]=\prod_{\ell<k}(x_k-x_\ell)$, we can rewrite the joint density~\eqref{complex:jpdf-cov} as
\begin{equation}
\rho_{N}^{\beta=2}(\lambda_1,\theta_1,\ldots,\lambda_N,\theta_N;t)=
\frac{1}{Z_{N,\nu}^{\beta=2}}\det_{1\leq k,\ell\leq N}\Big[e^{k(t\lambda_\ell+i\theta_\ell)}\Big]
\det_{1\leq k,\ell\leq N}\Big[e^{k(t\lambda_\ell-i\theta_\ell)}\Big]\prod_{n=1}^N t w_{\nu}^{\beta=2}(e^{t\lambda_n}).
\label{complex:jpdf-detdet}
\end{equation}
Here we have also absorbed the factor $\prod_{n}e^{2t\lambda_n}=\prod_{n}e^{t\lambda_n+i\theta_n}e^{t\lambda_n-i\theta_n}$ into the determinants. We may expand the first determinant as
\begin{equation}
\det_{1\leq k,\ell\leq N}[e^{k(t\lambda_\ell+i\theta_\ell)}]=\sum_{\sigma\in S_N} \sign\sigma \prod_{n=1}^N e^{n(t\lambda_{\sigma(n)}+i\theta_{\sigma(n)})},
\label{complex:expansion}
\end{equation}
where the sum is over all permutations and $\sign\sigma$ denotes the sign of the permutation. This expansion can be used to rewrite the joint density~\eqref{complex:jpdf-detdet} as a sum involving a single determinant. We expand the first determinant in~\eqref{complex:jpdf-detdet} and absorb the weight functions into the the second determinant. The signature, $\sign\sigma$, can then be absorbed into the determinant by reordering the rows in the determinant. Finally pulling the product from the expansion~\eqref{complex:expansion} into the determinant yields
\begin{equation}
\rho_{N}^{\beta=2}(\lambda_1,\theta_1,\ldots,\lambda_N,\theta_N;t)=
\frac{1}{Z_{N,\nu}^{\beta=2}}\sum_{\sigma\in S_N} 
\det_{1\leq k,\ell\leq N}\Big[e^{t(k+\ell)\lambda_{\sigma(\ell)}}e^{i(k-\ell)\theta_{\sigma(\ell)}} t\, w_{\nu}^{\beta=2}(e^{t\lambda_{\sigma(\ell)}})\Big]
\label{complex:jpdf-det}
\end{equation}
with normalization constant~\eqref{results:normal-complex} and the weight function~\eqref{results:weight}.

In order to simplify notation in the calculations below, we introduce the normalized function
\begin{equation}
f_{k\ell}^{\beta=2}(\lambda;t)\equiv \frac{2t\,\MeijerG{t}{0}{0}{t}{-}{\nu_1+\frac{k+\ell}{2},\ldots,\nu_t+\frac{k+\ell}{2}}{e^{2t\lambda}}}{\prod_{i=1}^t\Gamma\left[\nu_i+(k+\ell)/2\right]},\qquad
\int_{-\infty}^\infty d\lambda\, f_{k\ell}^{\beta=2}(\lambda;t)=1.
\label{complex:f_kl}
\end{equation}
Here the normalization follows directly from the integration formula~\eqref{meijer:meijer-moment}. Furthermore, it can be shown that $f_{k\ell}^{\beta=2}(\lambda;t)$ is real and non-negative, hence we may interpret it as a probability density. 

With definition~\eqref{complex:f_kl} in mind, we return to the joint density~\eqref{complex:jpdf-det}. If we write the weight as a Meijer $G$-function~\eqref{results:weight} and use the identity~\eqref{meijer:meijer-shift}, then we see that the first exponential in~\eqref{complex:jpdf-det} and the weight can be combined into a new Meijer $G$-function identical to the one appearing in~\eqref{complex:f_kl}. We insert the explicit expression for the normalization constant~\eqref{results:normal-complex} and after some standard manipulations we find
\begin{equation}
\rho_{N}^{\beta=2}(\lambda_1,\theta_1,\ldots,\lambda_N,\theta_N;t)=
\frac{1}{N!}\sum_{\sigma\in S_N} 
\det_{1\leq k,\ell\leq N}\bigg[\frac{e^{i(k-\ell)\theta_{\sigma(\ell)}} D_{k\ell}^{\beta=2}(t) f_{k\ell}^{\beta=2}(\lambda;t)}{2\pi}\bigg]
\label{complex:jpdf-final-det}
\end{equation}
with the coefficient $D_{k\ell}^{\beta=2}(t)$ defined by
\begin{equation}
D_{k\ell}^{\beta=2}(t)\equiv\prod_{i=1}^t\frac{\Gamma\left[\nu_i+(k+\ell)/2\right]}{\Gamma[\nu_i+k]^{1/2}\Gamma[\nu_i+\ell]^{1/2}}.
\label{complex:coefficient}
\end{equation}
So far no approximations have been used, hence the joint density~\eqref{complex:jpdf-final-det} is valid at any time $t$. In order to understand the large time asymptotics, we will look at $D_{k\ell}(t\gg1)$ and $f_{k\ell}^{\beta=2}(\lambda;t\gg 1)$ separately.

We start by evaluating the coefficient $D_{k\ell}(t)$. Recall that $\nu_i$ are non-negative integers, while $k$ and $\ell$ are positive integers. We note that
\begin{equation}
\Gamma[\nu_i+(k+\ell)/2]\leq \Gamma[\nu_i+k]^{1/2}\Gamma[\nu_i+\ell]^{1/2}
\label{complex:gamma-inequality}
\end{equation}
with equality if and only if $k=\ell$. This can be seen by writing the gamma functions as their integral representations and applying the Cauchy--Schwartz inequality. Comparing this inequality with the expression for the coefficient~\eqref{complex:coefficient}, we see that (recall that $\nu_\infty<\infty$)
\begin{equation}
D_{k\ell}^{\beta=2}(t)\longrightarrow \delta_{k\ell}\qquad \text{for}\qquad t\to\infty,
\label{complex:coefficient-limit}
\end{equation}
where $\delta_{k\ell}$ is a Kronecker delta. Note that $D_{kk}^{\beta=2}(t)=1$ even at finite time, while it can be verified that $D_{k\ell}^{\beta=2}(t)$ decays exponentially with time for $k\neq\ell$.

Similar to the discussion in~\cite{ABK:2014}, we will use the cumulant expansion to argue that the function $f_{k\ell}^{\beta=2}(\lambda;t)$ reduces to a Gaussian in the large time limit. In the following, we will interpret $f_{k\ell}^{\beta=2}(\lambda;t)$ as the probability density for a random variable $\lambda$. The cumulant-generating function $g_{k\ell}^{\beta=2}(\xi;t)$ is given by
\begin{equation}
g_{k\ell}^{\beta=2}(\xi;t)\equiv \log\Big( \int_{-\infty}^\infty d\lambda\, e^{\xi\lambda} f_{k\ell}^{\beta=2}(\lambda;t)\Big)
=\sum_{i=1}^t\log\frac{\Gamma[\nu_i+(k+\ell)/2+\xi/2t]}{\Gamma[\nu_i+(k+\ell)/2]}.
\end{equation}
Here we have used formula~\eqref{meijer:meijer-moment} to evaluate the integral. The cumulants are found by expanding the cumulant-generating function around $\xi=0$, hence the $n$-th cumulant is given by
\begin{equation}
\kappa_{k\ell,n}^{\beta=2}(t)\equiv \Big(\frac{\p}{\p\xi}\Big)^n g_{k\ell}^{\beta=2}(\xi;t)\Big\vert_{\xi=0}
=\sum_{i=1}^t\frac{1}{(2t)^n}\psi^{(n-1)}\Big(\nu_i+\frac{k+\ell}{2}\Big),
\end{equation}
where $\psi^{(n)}(x)$ is the $n$-th derivative of the digamma function also known as the polygamma function. As time tends to infinity the mean and the variance, i.e. the first and second cumulant, become
\begin{equation}
\mu_{k\ell}^{\beta=2}=\frac{1}{2}\average[\Big]{\psi\Big(\nu+\frac{k+\ell}{2}\Big)}
\qquad\text{and}\qquad
(\sigma_{k\ell}^{\beta=2})^2=\frac{1}{4t}\average[\Big]{\psi'\Big(\nu+\frac{k+\ell}{2}\Big)},
\label{complex:mean_kl}
\end{equation}
respectively. Recall that $k,\ell=1,\ldots,N$ and $\nu_t\to\nu_\infty<\infty$; this is enough to ensure that the averages are well-defined. To find the limiting distribution at large times, we switch from the random variable $\lambda$ to the standardized random variable $\widetilde \lambda\equiv (\lambda-\mu_{k\ell}^{\beta})/\sigma_{k\ell}^{\beta}$, which has mean zero and unit variance. The standardized cumulants are
\begin{equation}
\widetilde\kappa_{k\ell,1}^{\beta=2}(t)=0,\qquad
\widetilde\kappa_{k\ell,2}^{\beta=2}(t)=1 \qquad\text{and}\qquad
\widetilde\kappa_{k\ell,n}^{\beta=2}(t)\sim t^{1-n/2} \quad\text{for}\quad n\geq 3.
\end{equation}
We see that the higher order cumulants tend to zero at large times, and it follows from standard arguments that the limiting distribution is a Gaussian. Explicitly, we have
\begin{equation}
f_{k\ell}^{\beta=2}(\lambda;t\gg1)\approx\frac{1}{\sqrt{2\pi} \sigma_{k\ell}^{\beta=2}}\exp\bigg[-\frac{(\lambda-\mu_{k\ell}^{\beta=2})^2}{2(\sigma_{k\ell}^{\beta=2})^2}\bigg]\big(1+O(t^{-1/2})\big)
\label{complex:f-limit}
\end{equation}
with the mean and variance given by~\eqref{complex:mean_kl}.

We can now return to the joint density. Using the asymptotic behaviour~\eqref{complex:coefficient-limit} and~\eqref{complex:f-limit} in the joint density~\eqref{complex:jpdf-final-det} we see that
\begin{align}
\rho_{N}^{\beta=2}(\lambda_1,\theta_1,\ldots,\lambda_N,\theta_N;t\gg1)
=\frac{1}{N!}\sum_{\sigma\in S_N} \prod_{n=1}^N  \frac{1}{2\pi}f_{nn}^{\beta=2}(\lambda_{\sigma(n)};t\gg1),
\end{align}
where we recognize the right hand side as a permanent. If we introduce the notation $\mu_{n}^{\beta}\equiv\mu_{nn}^{\beta}$, $\sigma_{n}^{\beta}\equiv\sigma_{nn}^{\beta}$, and $f_{n}^{\beta}(\lambda;t)\equiv f_{nn}^{\beta}(\lambda;t)$, then we obtain the result~\eqref{results:permanent-2} stated in section~\ref{sec:results}. We see that the dependence on the phases, $\theta_n$, cancels out due to the Kronecker delta stemming from~\eqref{complex:coefficient-limit}, hence the integration over the phases can be performed trivially leaving the joint density for the Lyapunov exponents. Note that we could also obtain this result by first integrating over phases and then taking the large-$t$ limit. In fact, integration over the phases turns the joint density into a permanental point process even at finite times~\cite{Kostlan:1992,Rider:2004,AS:2013,AIS:2014},
\begin{equation}
\prod_{n=1}^N\int_{-\pi}^\pi d\theta_n\, \rho_{N}^{\beta=2}(\lambda_1,\theta_1,\ldots,\lambda_N,\theta_N;t)
=\frac{1}{N!}\perm_{1\leq k,\ell\leq N} \big[ f_{k}^{\beta=2}(\lambda_{\ell};t) \big].
\label{complex:jpdf-lyaponov}
\end{equation}
This means that the Lyapunov exponents defined by~\eqref{results:def-lyaponov-finite} are independent random variables at finite time; not only for $t\gg1$. In the large time limit the phases $\theta_n$ ($n=1,\ldots,N$) are independent random variables as well, but this independence does not extend to small times~\cite{Rider:2004}. 

\section{Derivation for quaternionic matrices}
\label{sec:quaternion}

After the discussion of complex matrices in the previous section, we are equipped to look at quaternionic matrices. We start at the joint probability density function for the eigenvalues~\eqref{results:jpdf-quaternion}. We use a standard identity for Vandermonde determinants~\cite{Mehta:2004},
\begin{equation}
\det_{\substack{k=1,\ldots,2N \\ \ell=1,\ldots,N}}\begin{bmatrix} z_\ell^{k-1} \\ z_\ell^{*k-1} \end{bmatrix}=
\prod_{1\leq k<\ell\leq N}\abs{z_k-z_\ell}^2\abs{z_k-z_\ell^*}^2\prod_{n=1}^N(z_n^*-z_n),
\end{equation}
and change variables from the eigenvalues, $z_n$, to the Lyapunov exponents, $\lambda_n$, and the phases, $\theta_n$. This leads to a joint density
\begin{equation}
\rho_{N}^{\beta=4}(\lambda_1,\theta_1,\ldots,\lambda_N,\theta_N;t)
=\frac{1}{Z_{N,\nu}^{\beta=4}}
\det_{\substack{k=1,\ldots,2N \\ \ell=1,\ldots,N}}\begin{bmatrix} e^{k(t\lambda_\ell+i\theta_\ell)} \\ e^{k(t\lambda_\ell-i\theta_\ell)} \end{bmatrix}
\prod_{n=1}^N2ite^{t\lambda_n}\sin\theta_n w_\nu^{\beta=4}(e^{t\lambda_n}).
\end{equation}
We expand the determinant and obtain
\begin{multline}
\rho_{N}^{\beta=4}(\lambda_1,\theta_1,\ldots,\lambda_N,\theta_N;t)= \\
\frac{1}{Z_{N,\nu}^{\beta=4}}\sum_{\sigma\in S_{2N}} \sign\sigma\prod_{n=1}^N2it\sin\theta_n e^{i(\sigma(n)-\sigma(n+N))\theta_n}e^{t(\sigma(n)+\sigma(n+N)+1)\lambda_n}w_\nu^{\beta=4}(e^{t\lambda_n}).
\label{quaternion:jpdf-expand}
\end{multline}
Analogously to the discussion of complex matrices in the previous section, we introduce a normalized, non-negative function,
\begin{equation}
f_{k\ell}^{\beta=4}(\lambda;t)\equiv \frac{2t\,\MeijerG{t}{0}{0}{t}{-}{2\nu_1+\frac{k+\ell+1}{2},\ldots,2\nu_t+\frac{k+\ell+1}{2}}{2^te^{2t\lambda}}}{\prod_{i=1}^t\Gamma\left[2\nu_i+(k+\ell+1)/2\right]},
\qquad  \int_{-\infty}^\infty d\lambda\, f_{k\ell}^{\beta=4}(\lambda;t)=1.
\label{quaternion:f_kl}
\end{equation}
With this definition we return to the joint density~\eqref{quaternion:jpdf-expand}. We recall the explicit form of the weight function~\eqref{results:weight} and use~\eqref{meijer:meijer-shift} to absorb the prefactor involving Lyapunov exponents into the Meijer $G$-function. Using~\eqref{quaternion:f_kl} and the explicit expression for the normalization constant~\eqref{results:normal-quaternion}, we write the joint density as  
\begin{multline}
\rho_{N}^{\beta=4}(\lambda_1,\theta_1,\ldots,\lambda_N,\theta_N;t)= \\
\frac{2^{N(N+1)t}}{\pi^NN!}\sum_{\sigma\in S_{2N}}\sign\sigma\, D_\sigma^{\beta=4}(t)\prod_{n=1}^N \frac{i\sin\theta_n e^{i(\sigma(n)-\sigma(n+N))\theta_n}}{2^{(\sigma(n)+\sigma(n+N)+1)t/2}}f_{\sigma(n),\sigma(n+N)}^{\beta=4}(\lambda_n),
\label{quaternion:jpdf-long}
\end{multline}
where the coefficient $D_\sigma^{\beta=4}(t)$ depends on the permutation $\sigma$,
\begin{equation}
D_\sigma^{\beta=4}(t)\equiv\prod_{i=1}^t\prod_{n=1}^N\frac{\Gamma[2\nu_i+(\sigma(n)+\sigma(n+N)+1)/2]}{\Gamma[2\nu_i+2n]}.
\label{quaternion:coefficient}
\end{equation}
As in the previous section we will discuss the asymptotic limit of the coefficient~\eqref{quaternion:coefficient} and the function~\eqref{quaternion:f_kl} separately, and then combine the results.

In order to evaluate the coefficient~\eqref{quaternion:coefficient} we will use the inequality~\eqref{complex:gamma-inequality} once again to write
\begin{equation}
\Gamma[2\nu_i+(k+\ell+1)/2]\leq (2\nu_i+\ell)^{1/2}\Gamma[2\nu_i+k]^{1/2}\Gamma[2\nu_i+\ell]^{1/2},
\end{equation}
where we have also used the recursive property $\Gamma[x+1]=x\Gamma[x]$. Note that $k$ and $\ell$ are interchangeable. Given a permutation, $\sigma\in S_{2N}$, it follows that 
\begin{equation}
\prod_{n=1}^N\Gamma[2\nu_i+(\sigma(n)+\sigma(n+N)+1)/2] 
\leq\prod_{n=1}^N(2\nu_i+\min\{\sigma(n),\sigma(n+N)\})^{1/2}\Gamma[2\nu_i+2n]^{1/2}\Gamma[2\nu_i+2n-1]^{1/2}.
\end{equation}
The maximal value of right hand side of this inequality only depends on the permutation through $\min\{\sigma(n),\sigma(n+N)\}$. It follows that the product takes its maximal value if and only if $\sigma(n)$ and $\sigma(n+N)$ comes in successive pairs, hence
\begin{equation}
\prod_{n=1}^N\Gamma[2\nu_i+(\sigma(n)+\sigma(n+N)+1)/2]\leq \prod_{n=1}^N\Gamma[2\nu_i+2n]
\end{equation}
with equality if and only if $\abs{\sigma(n)-\sigma(n+N)}=1$ for all $n$. Inserting this into the expression for the coefficient~\eqref{quaternion:coefficient}, we see that (recall that $\nu_\infty<\infty$)
\begin{equation}
D_\sigma^{\beta=4}(t)\longrightarrow \prod_{n=1}^N(\delta_{\sigma(n)+1,\sigma(n+N)}+\delta_{\sigma(n),\sigma(n+N)+1})
\qquad\text{for}\qquad
t\to\infty.
\label{quaternion:coefficient-limit}
\end{equation}
This implies that in the large time limit the sum over all permutations, $S_{2N}$, reduces to a sum over permutations of pairs, $S_N$. The other permutations will be exponentially suppressed at large times.

The evaluation of the asymptotic behaviour of the function $f_{k\ell}^{\beta=4}(\lambda;t)$ proceeds exactly as for complex ($\beta=2$) matrices and it follows that the limiting distribution is a Gaussian,
\begin{equation}
f_{k\ell}^{\beta=4}(\lambda;t\gg1)\approx\frac{1}{\sqrt{2\pi} \sigma_{k\ell}^{\beta=4}}\exp\bigg[-\frac{(\lambda-\mu_{k\ell}^{\beta=4})^2}{2(\sigma_{k\ell}^{\beta=4})^2}\bigg]\big(1+O(t^{-1/2})\big)
\label{quaternion:f-limit}
\end{equation}
with the mean and variance given by
\begin{equation}
\mu_{k\ell}^{\beta=4}=\frac12\log\frac12+\frac{1}{2}\average[\Big]{\psi\Big(2\nu+\frac{k+\ell+1}{2}\Big)}
\qquad\text{and}\qquad
(\sigma_{k\ell}^{\beta=4})^2=\frac{1}{4t}\average[\Big]{\psi'\Big(2\nu+\frac{k+\ell+1}{2}\Big)}.
\label{quaternion:mean_kl}
\end{equation}
The first term in the mean stems from the prefactor inside the argument of the Meijer $G$-function~\eqref{quaternion:f_kl}.

Combining the asymptotic behaviour for the coefficient~\eqref{quaternion:coefficient-limit} and the function~\eqref{quaternion:f-limit} in the expression for the joint density~\eqref{quaternion:jpdf-long} we find
\begin{equation}
\rho_{N}^{\beta=4}(\lambda_1,\theta_1,\ldots,\lambda_N,\theta_N;t\gg 1)=
\frac{1}{N!}\sum_{\sigma\in S_N}\prod_{n=1}^N \frac{2\sin^2\theta_n}{\pi}f_{2\sigma(n)-1,2\sigma(n)}^{\beta=4}(\lambda_n;t\gg1),
\end{equation}
where the right hand side can be written as a permanent. With the notation $\mu_{n}^{\beta}\equiv\mu_{2n-1,2n}^{\beta}$, $\sigma_{n}^{\beta}\equiv\sigma_{2n-1,2n}^{\beta}$, and $f_{n}^{\beta}(\lambda;t)\equiv f_{2n-1,2n}^{\beta}(\lambda;t)$, this becomes the result~\eqref{results:permanent-4} for quaternionic matrices stated in section~\ref{sec:results}. Unlike the complex case, the large time limit is not independent of the phases, $\theta_n$. This is due to the fact that the eigenvalues come in complex conjugate pairs which results in a repulsion from the real axis. However, the Lyapunov exponents are still independent Gaussian random variables. Moreover, if we integrate out the phases, then the Lyapunov exponents become independent random variables even at finite times~\cite{Rider:2004,AIS:2014}. Explicitly, we have
\begin{equation}
\prod_{n=1}^N\int_{0}^\pi d\theta_n\, \rho_{N}^{\beta=4}(\lambda_1,\theta_1,\ldots,\lambda_N,\theta_N;t)
=\frac{1}{N!}\perm_{1\leq k,\ell\leq N} \big[ f_{k}^{\beta=4}(\lambda_{\ell};t) \big],
\end{equation}
analogously to the expression for complex matrices~\eqref{complex:jpdf-lyaponov}.

\section{Derivation for real matrices}
\label{sec:real}

Finally, let us look at a product of real matrices with eigenvalues distributed according to~\eqref{results:jpdf-real}. Changing variables from the eigenvalues, $x_n$, to the Lyapunov exponents, $\lambda_n$, and the phases, $\theta_n$, the joint density~\eqref{results:jpdf-real} becomes
\begin{equation}
\rho_{N}^{\beta=1}(\lambda_1,\theta_1,\ldots,\lambda_N,\theta_N;t)
=\frac{1}{Z_{N,\nu}^{\beta=1}}\prod_{1\leq k<\ell\leq N}\abs{e^{t\lambda_k+i\theta_k}-e^{t\lambda_\ell+i\theta_\ell}}\prod_{n=1}^N te^{t\lambda_n}w_{\nu}^{\beta=1}(e^{t\lambda_n}).
\label{real:jpdf-cov}
\end{equation}
We recall that the phases $\theta_n$ only take the discrete values $0$ and $\pi$ because we only consider the eigenvalues which are located on the real axis. Our first step is to rewrite the joint density~\eqref{real:jpdf-cov} as a determinant,
\begin{equation}
\rho_{N}^{\beta=1}(\lambda_1,\theta_1,\ldots,\lambda_N,\theta_N;t)
=\frac{2^{-tN(N+1)/4}}{2^NN!}\abs[\bigg]{\det_{1\leq k,\ell\leq N}\bigg[\frac{e^{kt\lambda_\ell+i(k-1)\theta_\ell}2tw_{\nu}^{\beta=1}(e^{t\lambda_\ell})}{\prod_{i=1}^t\Gamma[(\nu_i+\ell)/2]}\bigg]}.
\label{real:jpdf-det}
\end{equation}
As in the two previous sections, we introduce a normalized and non-negative function
\begin{equation}
f_{k}^{\beta=1}(\lambda;t)\equiv \frac{2t\,\MeijerG{t}{0}{0}{t}{-}{\frac{\nu_1}{2}+\frac{k}{2},\ldots,\frac{\nu_t}{2}+\frac{k}{2}}{2^{-t}e^{2t\lambda}}}{\prod_{i=1}^t\Gamma\left[(\nu_i+k)/2\right]},\qquad
\int_{-\infty}^\infty d\lambda\, f_{k}^{\beta=1}(\lambda;t)=1.
\label{real:f_k}
\end{equation}
The joint density~\eqref{real:jpdf-det} simplifies considerably if we use the definition~\eqref{real:f_k} together with the explicit expression for the weight function~\eqref{results:weight} and the identity~\eqref{meijer:meijer-shift},
\begin{equation}
\rho_{N}^{\beta=1}(\lambda_1,\theta_1,\ldots,\lambda_N,\theta_N;t)
=\frac{1}{2^NN!}\abs[\Big]{\det_{1\leq k,\ell\leq N}\big[e^{i(k-1)\theta_\ell}f_{k}^{\beta=1}(\lambda_\ell;t)\big]}.
\label{real:jpdf-simple}
\end{equation}
Note that the Lyapunov exponents in the joint density~\eqref{real:jpdf-simple} are unordered. This can equivalently be expressed as a sum over all possible orderings, 
\begin{equation}
\rho_{N}^{\beta=1}(\lambda_1,\theta_1,\ldots,\lambda_N,\theta_N;t)
=\frac{1}{2^NN!}\sum_{\sigma\in S_N}\abs[\Big]{\det_{1\leq k,\ell\leq N}\big[e^{i(k-1)\theta_{\ell}}f_{k}^{\beta=1}(\lambda_{\ell};t)\big]}
\prod_{n=1}^{N-1}\Theta(\lambda_{\sigma(n+1)}-\lambda_{\sigma(n)}),
\end{equation}
where $\Theta(x)$ denotes the Heaviside function (the possibility that two Lyapunov exponents are equal has measure zero and may be disregarded).
If we write the determinant as a sum over permutation, then the joint density becomes
\begin{multline}
\rho_{N}^{\beta=1}(\lambda_1,\theta_1,\ldots,\lambda_N,\theta_N;t)=\\
\frac{1}{2^NN!}\sum_{\sigma\in S_N}\abs[\Bigg]{\sum_{\omega\in S_N}\sign\omega\prod_{k=1}^N e^{i(k-1)\theta_{\omega(k)}}f_{k}^{\beta=1}(\lambda_{\omega(k)};t)
\prod_{n=1}^{N-1}\Theta(\lambda_{\sigma(n+1)}-\lambda_{\sigma(n)})}.
\label{real:jpdf-long}
\end{multline}
So far no approximation have been used. 

At large times the normalized function~\eqref{real:f_k} can be evaluated using the same technique as in the two previous sections; we find that the limiting distribution is a Gaussian,
\begin{equation}
f_{k}^{\beta=1}(\lambda;t\gg1)\approx\frac{1}{\sqrt{2\pi} \sigma_{k}^{\beta=1}}\exp\bigg[-\frac{(\lambda-\mu_{k}^{\beta=1})^2}{2(\sigma_{k}^{\beta=1})^2}\bigg]\big(1+O(t^{-1/2})\big)
\label{real:f-limit}
\end{equation}
with the mean and variance given by
\begin{equation}
\mu_{k}^{\beta=1}=\frac12\log2+\frac{1}{2}\average[\Big]{\psi\Big(\frac{\nu+k}{2}\Big)}
\qquad\text{and}\qquad
(\sigma_{k}^{\beta=1})^2=\frac{1}{4t}\average[\Big]{\psi'\Big(\frac{\nu+k}{2}\Big)}.
\label{real:mean+var}
\end{equation}
In the strict $t\to\infty$ limit, the variances tends to zero and the Gaussians~\eqref{real:f-limit} become delta-peaks. Inserting this into~\eqref{real:jpdf-long}, we obtain
\begin{multline}
\rho_{N}^{\beta=1}(\lambda_1,\theta_1,\ldots,\lambda_N,\theta_N;t\to\infty)=\\
\frac{1}{2^NN!}\sum_{\sigma\in S_N}\abs[\Bigg]{\sum_{\omega\in S_N}\sign\omega\prod_{k=1}^N e^{i(k-1)\theta_{\omega(k)}}\delta(\lambda_{\omega(k)}-\mu_k^{\beta=1})
\prod_{n=1}^{N-1}\Theta(\lambda_{\sigma(n+1)}-\lambda_{\sigma(n)})}.
\end{multline}
We know from~\eqref{real:mean+var} that $\mu_k^{\beta=1}<\mu_\ell^{\beta=1}$ for $k<\ell$, hence the only contribution to the sum within the absolute value comes from the term $\omega=\sigma$ (all other terms are zero due the Heaviside functions). The absolute value cancels any overall sign of the remaining term, such that
\begin{equation}
\rho_{N}^{\beta=1}(\lambda_1,\theta_1,\ldots,\lambda_N,\theta_N;t\to\infty)
=\frac{1}{2^NN!}\sum_{\sigma\in S_N}\prod_{k=1}^N \delta(\lambda_{\sigma(k)}-\mu_k^{\beta=1}).
\label{real:lyaponov-delta}
\end{equation}
This is the statement that the finite-time Lyapunov exponents converge to the values~\eqref{results:lyaponov} as time tends to infinite.

At large but finite time we have the Gaussian approximation~\eqref{real:f-limit} rather than the delta-peaks. Note that the means of the Gaussians are independent of time while the variances decrease for increasing time~\eqref{real:mean+var}. It follows that at sufficiently large times, $f_{k}^{\beta=1}(\lambda;t\gg1)$ ($k=1,\ldots,N$) are well-separated Gaussian-peaks. Inserting this approximation into~\eqref{real:jpdf-long}, we see that at large times the dominant term in the sum inside the absolute value is $\omega=\sigma$, while all other terms will be exponentially suppressed due the Heaviside functions. The simplest illustration of this is when there is only one Heaviside function, i.e. only one constraint. In this case the Gaussian approximation~\eqref{real:f-limit} yields
\begin{equation}
\int_{-\infty}^\infty d\lambda_k \int_{-\infty}^\infty d\lambda_\ell\, f_{\ell}^{\beta=1}(\lambda_k;t\gg1) f_{k}^{\beta=1}(\lambda_\ell;t\gg1)\Theta(\lambda_k-\lambda_\ell)
\approx\frac12\erfc\Bigg[\frac{\mu_k^{\beta=1}-\mu_\ell^{\beta=1}}{\big(2(\sigma_k^{\beta=1})^2+2(\sigma_\ell^{\beta=1})^2\big)^{1/2}}\Bigg].
\end{equation}
This integral gives the probability for a particular ordering of Lyapunov exponents and we see that it is exponentially close to unity for $k<\ell$ and exponentially suppressed for $k>\ell$. It is clear that the idea of this argument can be extended to the general case, and it follows that at large but finite time~\eqref{real:lyaponov-delta} is replaced by
\begin{equation}
\rho_{N}^{\beta=1}(\lambda_1,\theta_1,\ldots,\lambda_N,\theta_N;t\gg1)
=\frac{1}{2^NN!}\sum_{\sigma\in S_N}\prod_{k=1}^Nf_{k}^{\beta=1}(\lambda_{\sigma(k)};t\gg1).
\end{equation}
Here, we recognize the right hand side as the permanent from the expression~\eqref{results:permanent-1} stated in section~\ref{sec:results}. Note that integrating out the Lyapunov exponents and summing over the phases yields unity which confirms that all eigenvalues are real with probability equal to unity. This is consistent with the striking property previously observed in~\cite{Forrester:2014a,Lakshminarayan:2013}.

\section{Conclusions and outlook}
\label{sec:conclusion}

In this paper we have studied the Lyapunov exponents constructed from the eigenvalues of a product of independent real, complex, or quaternionic Ginibre matrices. We have seen that all three classes are solvable as the number of matrices tends to infinity. If we consider a product of a large number of finite size matrices, then the Lyapunov exponents become independent Gaussian random variables, which conveniently can be expressed as a permanental point process. This implicitly implies that the moduli of the eigenvalues are independent and log-normal distributed. The results are in agreement with known results for Lyapunov exponents~\cite{Newman:1986,Forrester:2013,Kargin:2014} as well as a previous result regarding products of complex Ginibre matrices~\cite{ABK:2014}. The properties of the phases of the eigenvalues have a striking dependence on whether real, complex or quaternionic matrices are considered. For real ($\beta=1$) matrices all eigenvalues are real, for complex ($\beta=2$) matrices the 
eigenvalue distribution is 
rotational invariant in the complex plane, while the eigenvalues of quaternionic ($\beta=4$) matrices have a sine squared repulsion from the real axis. The rotational invariance of the eigenvalue spectrum for $\beta=2$ is inherited from rotational invariance of the weight function and is valid for an arbitrary number of matrices of any size, while the angular dependence for $\beta=1$ and $\beta=4$ are inherently connected to the asymptotic limit. Note that the angular structure for $\beta=1$ reaffirms a previous result~\cite{Lakshminarayan:2013,Forrester:2014a}, while the $\beta=4$ case confirms a conjecture stated in~\cite{ABK:2014}.

The discussion in this paper has been limited to Ginibre matrices, but it is tempting to extend the discussion to other ensembles where explicit results are known, e.g. products of truncated unitary matrices~\cite{ARRS:2013,IK:2014,ABKN:2014}. Moreover, it would be interesting to consider double scaling limits where both the number of matrices and the matrix dimensions tend to infinity. It was shown in~\cite{ABK:2014} that when we consider the global density then the two limits are interchangeable. In fact, the global density for Lyapunov exponents is a universal distribution known as the triangular law~\cite{Newman:1986,IN:1992}. However, the interchangeability of the two limits is by no means guaranteed on the local scale. Comparing the results presented in this paper and in~\cite{ABK:2014} with previous results where the matrix dimensions rather than the number matrices is taken to infinity~\cite{AB:2012,Ipsen:2013,IK:2014,LW:2014} (see also~\cite{Forrester:2014a}) suggest that non-trivial behaviour may 
be found in such double scaling limits. Actually, a very recent paper~\cite{JQ:2014} has observed a crossover for the largest moduli of an eigenvalue from the Gumbel distribution (Ginibre-like universality) to a log-normal distribution (Lyapunov-like universality). 

Another intriguing question is to ask about the behaviour of the eigen- and singular values as a function of the number of matrices for a single realization of the product matrix. This generates a stochastic motion in discrete time (given by the number of matrices) of a set of interacting particles (given by the eigen- or singular values). We have seen that the fluctuations between different realizations of the product matrix give rise to the same structure for both eigen- and singular values, but we have also seen that numerical simulations suggest that fluctuations in time differ. To our knowledge, this difference has yet to be discussed in the literature.

\paragraph{Acknowledgement} G. Akemann and M. Kieburg are thanked for useful discussions and for comments on the first draft of this paper. The author is supported by the German Science Foundation (DFG) through the International Graduate College `Stochastics and Real World Models' (IRTG 1132).

\appendix

\section{Functional identities}
\label{sec:meijer}
This appendix gives the definition of the Meijer $G$-function and states two identities (these can also be found in~\cite{GR:2000} but are repeated here for easy reference).
The Meijer $G$-function is defined by a contour integral in the complex plane,
\begin{equation}
\MeijerG{m}{n}{p}{q}{a_1,\,\ldots\,,\,a_p}{b_1,\,\ldots\,,\,b_q}{z}\equiv
\frac{1}{2\pi i}\int_L du\,z^u\frac{\prod_{i=1}^m\Gamma[b_i-u]\prod_{i=1}^n\Gamma[1-a_i+u]}{\prod_{i=n+1}^p\Gamma[a_i-u]\prod_{i=m+1}^q\Gamma[1-b_i+u]},
\label{meijer:meijer-def}
\end{equation}
where the contour $L$ runs from $-i\infty$ to $+i\infty$ and is chosen such that it separates the poles stemming from $\Gamma[b_i-u]$ and the poles stemming from $\Gamma[1-a_i+u]$. In this paper we use two properties of the Meijer $G$-function: (i) Multiplication by a power can be absorbed into the Meijer $G$-function as a shift in the parameters,
\begin{equation}
z^c\, \MeijerG{m}{n}{p}{q}{a_1,\ldots,a_p}{b_1,\ldots,b_q}{z}=\MeijerG{m}{n}{p}{q}{a_1+c,\ldots,a_p+c}{b_1+c,\ldots,b_q+c}{z},
\label{meijer:meijer-shift}
\end{equation}
and (ii) the moments of a Meijer $G$-function can be expressed as a ratio of gamma functions,
\begin{equation}
\int_0^\infty dr\,r^{s-1}\,\MeijerG{m}{n}{p}{q}{a_1,\ldots,a_p}{b_1,\ldots,b_q}{z r}=
\frac{1}{z^s}\,\frac{\prod_{i=1}^m\Gamma[b_i+s]\prod_{j=1}^n\Gamma[1-a_j-s]}{\prod_{k=m+1}^q\Gamma[1-b_k-s]\prod_{\ell=n+1}^p\Gamma[a_\ell+s]}.
\label{meijer:meijer-moment}
\end{equation}


\raggedright

\bibliographystyle{h-physrev}
\bibliography{ref}

\begin{thebibliography}{10}

\bibitem{CPV:1993}
A.~Crisanti, G.~Paladin, and A.~Vulpiani,
\newblock {\em Products of random matrices} (Springer, 1993).

\bibitem{Beenakker:1997}
C.~W.~J. Beenakker,
\newblock Rev. Mod. Phys. {\bf 69}, 731 (1997), arXiv:cond-mat/9612179.

\bibitem{Haake:2010}
F.~Haake,
\newblock {\em Quantum Signatures of Chaos, 3rd ed.} (Springer, 2010).

\bibitem{Osborn:2004}
J.~C. Osborn,
\newblock Phys. Rev. Lett. {\bf 93}, 222001 (2004), arXiv:hep-th/0403131.

\bibitem{Akemann:2007}
G.~Akemann,
\newblock Int. J. Mod. Phys. A {\bf 22}, 1077 (2007), arXiv:hep-th/0701175.

\bibitem{Caswell:2001}
H.~Caswell,
\newblock {\em Matrix population models, 2nd ed.} (Sinauer Associates, 2001).

\bibitem{Muller:2002}
R.~R. Muller,
\newblock IEEE Trans. Inf. Theor. {\bf 48}, 2086 (2006).

\bibitem{TV:2004}
A.~Tulino and S.~Verd{\'u},
\newblock {\em Random Matrix Theory And Wireless Communications} (Now, 2004).

\bibitem{NS:2006}
A.~Nica and R.~Speicher,
\newblock {\em Lectures on the Combinatorics of Free Probability} (Cambridge
  University Press, 2006).

\bibitem{Benettin:1984}
G.~Benettin,
\newblock Physica D {\bf 13}, 211 (1984).

\bibitem{PV:1986}
G.~Paladin and A.~Vulpiani,
\newblock J. Phys. A {\bf 19}, 1881 (1986).

\bibitem{DVP:1978}
B.~Derrida, J.~Vannimenus, and Y.~Pomeau,
\newblock J. Phys. C {\bf 11}, 4749 (1978).

\bibitem{FK:1960}
H.~Furstenberg and H.~Kesten,
\newblock Ann. Math. Stat. {\bf 31}, 457 (1960).

\bibitem{Oseledec:1968}
V.~I. Oseledec,
\newblock Trans. Moscow Math. Soc. {\bf 19}, 197 (1968).

\bibitem{Raghunathan:1979}
M.~S. Raghunathan,
\newblock Israel J. Math. {\bf 32}, 356 (1979).

\bibitem{JLJN:2002}
A.~D. Jackson, B.~Lautrup, P.~Johansen, and M.~Nielsen,
\newblock Phys. Rev. E {\bf 66}, 066124 (2002), arXiv:physics/0202037.

\bibitem{Newman:1986}
C.~M. Newman,
\newblock Commun. Math. Phys. {\bf 103}, 121 (1986).

\bibitem{Forrester:2013}
P.~J. Forrester,
\newblock J. Stat. Phys. {\bf 151}, 796 (2013), arXiv:1206.2001.

\bibitem{Kargin:2014}
V.~Kargin,
\newblock J. Stat. Phys. {\bf 157}, 70 (2014), arXiv:1306.6576.

\bibitem{AB:2012}
G.~Akemann and Z.~Burda,
\newblock J. Phys. A {\bf 45}, 465201 (2012), arXiv:1208.0187.

\bibitem{AKW:2013}
G.~Akemann, M.~Kieburg, and L.~Wei,
\newblock J. Phys. A {\bf 46}, 275205 (2013), arXiv:1303.5694.

\bibitem{AIK:2013}
G.~Akemann, J.~R. Ipsen, and M.~Kieburg,
\newblock Phys. Rev. E {\bf 88}, 052118 (2013), arXiv:1307.7560.

\bibitem{Ipsen:2013}
J.~R. Ipsen,
\newblock J. Phys. A {\bf 46}, 265201 (2013), arXiv:1301.3343.

\bibitem{Forrester:2014a}
P.~J. Forrester,
\newblock J. Phys. A {\bf 47}, 065202 (2014), arXiv:1309.7736.

\bibitem{IK:2014}
J.~R. Ipsen and M.~Kieburg,
\newblock Phys. Rev. E {\bf 89}, 032106 (2014), arXiv:1310.4154.

\bibitem{LS:1991}
N.~Lehmann and H.-J. Sommers,
\newblock Phys. Rev. Lett. {\bf 67}, 941 (1991).

\bibitem{Edelman:1997}
A.~Edelman,
\newblock J. Multivariate Anal. {\bf 60}, 203 (1997).

\bibitem{APS:2009}
G.~Akemann, M.~J. Phillips, and H.-J. Sommers,
\newblock J. Phys. A {\bf 42}, 012001 (2009), arXiv:0810.1458.

\bibitem{APS:2010}
G.~Akemann, M.~J. Phillips, and H.-J. Sommers,
\newblock J. Phys. A {\bf 43}, 085211 (2010), arXiv:0911.1276.

\bibitem{Lakshminarayan:2013}
A.~Lakshminarayan,
\newblock J. Phys. A {\bf 46}, 152003 (2013), arXiv:1301.7601.

\bibitem{ABK:2014}
G.~Akemann, Z.~Burda, and M.~Kieburg,
\newblock J. Phys. A {\bf 47}, 395202 (2014), arXiv:1406.0803.

\bibitem{GA:1970}
M.~R. Gardner and W.~R. Ashby,
\newblock Nature {\bf 228}, 784 (1970).

\bibitem{May:1972}
R.~M. May,
\newblock Nature {\bf 238}, 413 (1972).

\bibitem{Mehta:2004}
M.~L. Mehta,
\newblock {\em Random Matrices, 3rd ed.} (Elsevier Science, 2004).

\bibitem{ARRS:2013}
K.~Adhikari, N.~K. Reddy, T.~R. Reddy, and K.~Saha,
\newblock preprint  (2013), arXiv:1308.6817.

\bibitem{GSO:1987}
I.~Goldhirsch, P.-I. Sulem, and S.~A. Orszag,
\newblock Physica D {\bf 27}, 311 (1987).

\bibitem{Forrester:2015}
P.~J. Forrester,
\newblock arXiv:1501.05702  (2015).

\bibitem{Kostlan:1992}
E.~Kostlan,
\newblock Lin. Alg. Appl. {\bf 162–164}, 385 (1992).

\bibitem{Rider:2004}
B.~Rider,
\newblock Probab. Theory Rel. {\bf 130}, 337 (2004), arXiv:math/0312043.

\bibitem{AS:2013}
G.~Akemann and E.~Strahov,
\newblock J. Stat. Phys. {\bf 151}, 987 (2013), arXiv:1211.1576.

\bibitem{AIS:2014}
G.~Akemann, J.~R. Ipsen, and E.~Strahov,
\newblock Random Matrices {\bf 03}, 1450014 (2014), arXiv:1404.4583.

\bibitem{ABKN:2014}
G.~Akemann, Z.~Burda, M.~Kieburg, and T.~Nagao,
\newblock J. Phys. A {\bf 47}, 255202 (2014), arXiv:arXiv:1310.6395.

\bibitem{IN:1992}
M.~Isopi and C.~M. Newman,
\newblock Comm. Math. Phys. {\bf 143}, 591 (1992).

\bibitem{LW:2014}
D.-Z. Liu and Y.~Wang,
\newblock arXiv:1411.2787  (2014).

\bibitem{JQ:2014}
T.~Jiang and Y.~Qi,
\newblock arXiv:1411.1833  (2014).

\bibitem{GR:2000}
I.~S. Gradshteyn and I.~M. Ryzhik,
\newblock {\em Table of Integrals, Series, and Products, 7th ed.} (Academic
  Press, 2007).

\end{thebibliography}

\end{document}